\def\pheq{\phantom{=}}
\newcommand{\Deqn}[1]{{Eq.~(\ref{#1})}}
\newcommand{\Deqns}[1]{{Eqs.~(\ref{#1})}}

\newcommand{\Dfig}[1]{{Fig.~\ref{#1}}}
\newcommand{\beq}{\begin{equation}}
\newcommand{\eeq}{\end{equation}}
\newcommand{\bea}{\begin{eqnarray}}
\newcommand{\eea}{\end{eqnarray}}
\newcommand{\cn}{\mathop{\mathrm{cn}}\nolimits}

\documentclass[twocolumn,prd,aps,amssymb,nofootinbib,epsf]{revtex4}
\usepackage{bm}
\usepackage{amsmath}
\usepackage{amsfonts}
\usepackage{graphicx}

\begin{document}

\title{Stochastic Backgrounds of Gravitational Waves from Cosmological 
Sources: Techniques and Applications to Preheating}
\author{Larry R. Price}
\email{larry@gravity.phys.uwm.edu}
\author{Xavier Siemens}
\email{siemens@gravity.phys.uwm.edu}
\affiliation{Center for Gravitation and Cosmology, Department of Physics, 
University of Wisconsin--Milwaukee, P.O. Box 413, Milwaukee, Wisconsin 53201, 
USA}
\begin{abstract}
  Several mechanisms exist for generating a stochastic background of
  gravitational waves in the period following inflation. These
  mechanisms are generally classical in nature, with the gravitational
  waves being produced from inhomogeneities in the fields that
  populate the early universe and not quantum fluctuations. The
  resulting stochastic background could be accessible to next
  generation gravitational wave detectors.  We develop a framework for
  computing such a background analytically and computationally.  As an
  application of our framework, we consider the stochastic background
  of gravitational waves generated in a simple model of preheating.
\end{abstract}

\maketitle

\section{\label{intro}Introduction}

Prior to the time of recombination the universe is opaque to
electromagnetic waves. The detection of a gravitational wave
background generated before recombination would provide a unique
observational window. Observations of this background would afford us
rare and powerful probes of early universe physics and cosmology and
could have profound implications.

The detection and study of gravitational waves is currently at the
forefront of fundamental physics research. A world-wide network of
gravitational wave detectors is poised to make the first detection.
The Laser Interferometer Gravitational-Wave Observatory (LIGO) and the
Virgo interferometer have already achieved unprecedented sensitivities
at frequencies around $100$~Hz and are currently undergoing upgrades.
Future detectors such as Advanced LIGO and the Laser Interferometer
Space Antenna (LISA) promise to provide even more sensitivity and
allow for observations at other frequencies.

Inflation produces a stochastic background of gravitational waves
through the amplification of primordial quantum
fluctuations~\cite{Starobinsky:1979ty,Allen:1987bk}.  Unfortunately 
this background is too weak to be directly detected with existing
instruments. It might, however, be observed via B-modes in the cosmic 
microwave background
(CMB)~\cite{Spergel:2006hy,Peiris:2006sj}. In the period following
inflation there are a number of mechanisms that could result in the
production of an additional gravitational wave background; for example
phase transitions~\cite{Kosowsky:1991ua,Kosowsky:1992rz,Kosowsky:1992vn,
Krauss:1991qu,Kamionkowski:1993fg,Felder:2000hj,Kosowsky:2001xp}, as well as 
reheating and preheating~\cite{Kofman:1994rk,Khlebnikov:1997di,Greene:1997fu,
  Kofman:1997yn,GarciaBellido:1997wm,GarciaBellido:1998gm,
  Easther:2006gt,Easther:2006vd,GarciaBellido:2007af,
  GarciaBellido:2007dg,Dufaux:2007pt,Easther:2007vj,Easther:2008sx,
Caprini:2007xq,JonesSmith:2007ne}.
These mechanisms are generally classical in nature: The gravitational
waves are generated through inhomogeneities in the various matter
fields that populate the early universe.  The gravitational waves
produced by some of these mechanisms are close to being detectable by
Advanced LIGO~\cite{GarciaBellido:1998gm, Easther:2006vd}, and well
within the reach of LISA~\cite{Hogan:2006va}.

In many of these models the effects of the expansion of the universe
cannot be neglected. The gravitational waves can be produced on time and length
scales comparable to the Hubble scale.  In this paper we develop a
framework for computing gravitational wave backgrounds in such
situations and apply it to a simple well-studied model of preheating.

The realization that preheating could lead to the efficient production
of a gravitational wave background was first made by Khlebnikov and
Tkachev~\cite{Khlebnikov:1997di}.  They made an analytic
estimate of the stochastic background produced by preheating in a
quartic inflation model.  To estimate the stochastic background they
used a flat space formula due to Weinberg~\cite{137} for the energy in
gravitational waves per unit solid angle. They
found relatively large values of the background
peaked at frequencies around $f \sim 10^8$~Hz. 
A similar estimate by Garcia-Bellido~\cite{GarciaBellido:1998gm},
showed that in hybrid inflation the frequency of the peak in the
spectrum could be brought down to the 1~kHz range where Advanced LIGO
might detect it. Almost a decade later, the problem was independently revisted by  
Garcia-Bellido and Figueroa~\cite{GarciaBellido:2007dg} and Easther and
Lim~\cite{Easther:2006vd}. The former~\cite{GarciaBellido:2007dg} studied hybrid 
inflation models using a flat space code that evolved the scalar fields as well 
as the metric perturbation. The latter~\cite{Easther:2006vd} evolved the scalar 
fields in a Friedmann-Robertson-Walker
universe using Felder and Tkachev's {\sc
  LatticeEasy}~\cite{Felder:2000hq}, considered quartic as well as
quadratic inflationary potentials, and used Weinberg's flat space
formula. Easther, Giblin, and Lim~\cite{Easther:2006vd,Easther:2007vj}
later developed a new computational strategy to include the effects of
cosmological expansion. They coupled {\sc
  LatticeEasy} to an integrator for the equation for the metric
perturbation (using a mode decomposition) which they used to study
hybrid inflation models. They found that the amplitude of the
background is energy scale independent, and confirmed the analytic
estimates of Garcia-Bellido~\cite{GarciaBellido:1998gm} showing hybrid
inflation models might lead to a signal detectable by Advanced LIGO.
Later, Garcia-Bellido, Figueroa, and
Sastre~\cite{GarciaBellido:2007af} re-examined the problem in an expanding
spacetime using {\sc LatticeEasy} and a configuration space integrator for the 
metric perturbation.  In the meantime
Dufaux, Bergman, Felder, Kofman, and Uzan~\cite{Dufaux:2007pt} studied
the problem using Green's functions. They constructed an approximate
Green's function valid for gravitational waves generated well inside
the horizon for any kind of cosmological expansion.  The methods we 
develop in this paper are most similar to this work.  

In Section~\ref{framew}, using a mode decomposition in the equation for the 
evolution of the metric perturbation, we construct {\emph{exact}} Green's
functions for radiation- and matter-dominated expansions.  Since our
Green's functions are exact we faithfully recover the perturbations at
all wavelengths. We are, however, tied to particular types of
expansion (radiation- or matter-dominated).  We use these Green's
functions to construct an analytic expression for the energy emitted
in gravitational waves as a function the Fourier transform of the
stress energy tensor (in conformal coordinates) analogous to
Weinberg's flat space formula~\cite{137}. This expression is useful
when the evolution of the fields is known as well as for analytic
estimates.  We also use these Green's functions to write down the energy
density in gravitational waves in a form suitable for lattice simulations.
In Section~\ref{example} we introduce the simple model of preheating that we
use to validate our framework.  This is followed by a prescription of how to 
translate the results of our simulations to values today.  Finally we present 
and discuss our results.  Our conclusions are found in Section~\ref{conclusions}.

\section{\label{framew}Theoretical framework}

In this section we will lay out the basic framework for computing
stochastic backgrounds given some matter source ($T_{\mu\nu}$).  We
will start by solving the perturbed Einstein equations for radiation-
and matter-dominated expansion using exact Green's functions in
conformal coordinates. In Appendix~\ref{mpct} we provide these Green's
functions in comoving coordinates.  The metric perturbation determines
the effective stress-energy tensor for gravitational waves, which, in
turn, leads to an expression for the energy density in gravitational
waves.  We will develop this in both analytic and computational
directions.  On the analytic side we will provide a Weinberg-like
formula for the energy in gravitational waves as a function of solid
angle valid for expanding spacetimes.  On the computational side, we
will write down the same equations in a form useful for lattice
simulations.

\subsection{Metric perturbations: Conformal Coordinates} 

We will be primarily working in a spatially flat Friedmann-Robertson-Walker (FRW)
metric in conformal coordinates:
\beq
ds^2=a^2(\eta)(-d\eta^2+dx^2+dy^2+dz^2).
\eeq
Our interest is in the first order metric perturbation, $h_{\mu\nu}$.
We choose to work in a spatial transverse-traceless 
gauge~\cite{Misner:1974qy} defined by
\begin{eqnarray}
h_{\mu 0}^{\rm TT} &=&0,\label{gc1}\\
\partial_i h_{ij}^{\rm TT} &=&0,\label{gc2}\\
h_{ii}^{\rm TT}&=&0,\label{gc3}
\end{eqnarray}
where $\mu$ is a spacetime index, $i,j=1,2,3$ labels spatial indices and
$\partial_i =\frac{\partial}{\partial x^i}$.  The physical metric is
then
\beq
ds^2=a^2(\eta)[-d\eta^2+(\delta_{ij}+h_{ij}^{\rm TT})dx^idx^j],
\label{physmet}
\eeq
and the perturbed Einstein equations take on a particularly simple
form:
\beq
\ddot{h}^{\rm TT}_{ij} + 2\frac{\dot{a}(\eta)}{a(\eta)}
\dot{h}^{\rm TT}_{ij} -\nabla^2h^{\rm TT}_{ij} = 16\pi T_{ij}^{\rm
  TT},
\label{eqm}
\eeq
where the overdot denotes the derivative with respect to conformal
time, $\nabla^2$ is the three dimensional Laplacian in Euclidean space and the 
TT superscript denotes the transverse-traceless projection described 
below.

For a generic perturbation of an arbitrary spacetime there is an
inherent ambiguity in differentiating between the background and the
perturbation.  The ambiguity is essentially the gauge choice one makes
when identifying points on the background spacetime with those on the
physical (perturbed) spacetime~\cite{Stewart:1974uz}. For a spatially
flat FRW background, where the expansion is driven by a perfect fluid
source with comoving velocity $u^\mu$, the full (``background +
perturbation") stress-energy tensor is
\beq
T_{\mu\nu}= (\rho +p) u_\mu u_\nu + p g_{\mu\nu} + \pi_{\mu\nu},\label{fullsource}
\eeq
where $\rho$ and $p$ are the energy 
density and pressure, respectively, $g_{\mu\nu}$ is the metric of the 
background (FRW) spacetime and $\pi_{\mu\nu}=\pi_{\nu\mu}$ is the 
anisotropic stress\footnote{Some authors write $\pi_{\mu\nu}=a^2\Pi_{\mu\nu}$ in 
FRW backgrounds.}. We consider the contribution of the anisotropic stress to be
a perturbation of the otherwise homogeneous and isotropic background (with the 
expansion sourced by the first two terms in \Deqn{fullsource}).  The anisotropic
stress tensor satisfies $u^\mu\pi_{\mu\nu}=\pi^\mu{}_\mu=0$, so it is both 
purely spatial and traceless.  In this situation it is clear that the only 
surviving term under the (spatial) transverse-traceless projection is 
$\pi_{ij}^{\rm TT}$, so this choice of gauge provides an 
unambiguous splitting of the (homogeneous) background and the perturbation.  
For this paper we will be working exclusively in the TT gauge and will 
thus simply write our source term as $T_{ij}^{\rm TT}$.

Solving \Deqn{eqm} is easiest if we first perform a spatial Fourier
transform while leaving the dependence on conformal time unchanged.  Our
convention for the Fourier transform is given by
\beq
f(\eta,\mathbf{x}) = \frac{1}{(2\pi)^2}\int_{-\infty}^\infty d\nu\,
d^3\mathbf{k}\, e^{-i(\nu \eta -\mathbf{k}\cdot\mathbf{x})} f(\nu,\mathbf{k}).
\label{ftcon}
\eeq 
The perturbed Einstein equations then become
\beq
\ddot{h}^{\rm TT}_{ij}(\eta,\mathbf{k}) 
+ 2\frac{\dot{a}(\eta)}{a(\eta)}\dot{h}^{\rm TT}_{ij}(\eta,\mathbf{k})
+ \omega^2h^{\rm TT}_{ij}(\eta,\mathbf{k})  = 16\pi T_{ij}^{\rm
  TT}(\eta,\mathbf{k}) ,
\label{eqmsft}
\eeq
with $\omega^2=\mathbf{k}\cdot\mathbf{k}$.  

The transverse-traceless projection is easily described in momentum
space.  The TT part of some spatial tensor, $T_{ij}$, is given 
by~\cite{Misner:1974qy}
\beq
T_{ij}^{\rm TT}(\mathbf{k}) = \{P_{im}(\mathbf{k})P_{jn}(\mathbf{k})
-\frac{1}{2}P_{ij}(\mathbf{k})P_{mn}(\mathbf{k})\}T_{mn}(\mathbf{k}),
\label{TTT}
\eeq
where
\beq
P_{ij}(\mathbf{k})=\delta_{ij}-\frac{k_ik_j}{\omega^2}.
\label{proj}
\eeq

In order to solve for the metric perturbation, we must know something about
the evolution of the scale factor.  We will restrict our attention to simple power 
law evolution of the form 
\beq
a(\eta)=\alpha\eta^{n},
\eeq
for some constants $\alpha$ and $n$. The solution in terms of a Green's function is given 
quite simply by
\bea
h_{ij}^{\rm
  TT}(\eta,\mathbf{k})&=&16\pi\int_{\eta_i}^{\eta_<}d\eta'\,
\omega\frac{\eta^{\prime\,n+1}}{\eta^{n-1}} \Bigl[j_{n-1}(\omega\eta')y_{n-1}(\omega\eta)\nonumber\\
&\pheq&\phantom{\pi\omega\int_{\eta_i}^{\eta_<}d\eta}
-j_{n-1}(\omega\eta)y_{n-1}(\omega\eta')\Bigr]
T_{ij}^{\rm TT}(\eta',\mathbf{k}),\nonumber\\
\label{gengf}
\eea
where $j_n(z)$ and $y_n(z)$ are spherical Bessel functions of the first and 
second kind, respectively. Our primary interest is in the particular cases
\beq
a(\eta)=\alpha\eta\qquad{\rm and} \qquad a(\eta)=\alpha\eta^2,
\eeq
which correspond to radiation- and matter-dominated expansion, respectively. The 
general solution in \Deqn{gengf} then becomes  
\beq
h_{ij}^{\rm
  TT}(\eta,\mathbf{k})=\frac{16\pi}{\omega\eta}\int_{\eta_i}^{\eta_<}d\eta'\,
\eta' \sin[\omega(\eta-\eta')]T_{ij}^{\rm TT}(\eta',\mathbf{k}),\label{eqhab}
\eeq
for radiation-dominated expansion and
\bea
 h_{ij}^{\rm TT}(\eta,\mathbf{k})
&=&\frac{16\pi}{(\omega\eta)^3}\int_{\eta_i}^{\eta_<}d\eta'\, 
\eta' \Bigl\{(1+\omega^2\eta\eta')\sin[\omega(\eta-\eta')]
\nonumber
\\
&\pheq&\phantom{\frac{16\pi}{(\omega\eta)^3}\int_{\eta_i}^{\eta}d\eta'\,}
- \omega(\eta-\eta')\cos[\omega(\eta-\eta')]\Bigr\}
\nonumber
\\
&\pheq&\phantom{\frac{16\pi}{(\omega\eta)^3}\int_{\eta_i}^{\eta}d\eta'\,}
\times T_{ij}^{\rm TT}(\eta',\mathbf{k}),
\label{eqhabmattdom} 
\eea 
for matter-dominated expansion. The lower limit of integration in both
\Deqns{eqhab} and (\ref{eqhabmattdom}) is determined by the time the
source turns on, $\eta_i$. The upper limit is $\eta_<\equiv \min(\eta,\eta_f)$,
where $\eta$ is the time at which the metric perturbation is evaluated and 
$\eta_f$ is the time the source turns off.

It is interesting to note that both radiation- and matter-dominated
expansion lead to relatively simple Green's functions.
This is a result of the fact that we are working in momentum space.  In
configuration space the Green's function for radiation-dominated
expansion only has support on the light-cone. The matter-dominated
expansion Green's function contains a ``tail" term that has support
{\it inside} the past lightcone~\cite{Caldwell:1993xw}.  This severely
complicates a configuration space calculation of the metric
perturbation.

\subsection{\label{gravrad}Energy density in gravitational radiation}
 
The effective stress-energy tensor for gravitational radiation is given 
covariantly (in a transverse-traceless gauge) by~\cite{Misner:1974qy}
\beq
T_{\mu\nu}^{\rm gw} =\frac{1}{32\pi}\Bigl\langle\nabla_\mu 
\gamma_{\alpha\beta}^{\rm TT}\nabla_\nu \gamma^{\alpha\beta\,{\rm
    TT}}\Bigr\rangle,
\label{Tabavg}
\eeq 
where $\gamma_{\mu\nu}$ is the metric perturbation 
($g_{\mu\nu}=g_{\mu\nu}^{\rm background} +\gamma_{\mu\nu}$) and
the angle brackets denote a spatial average over several wavelengths
and the covariant derivative is compatible with the background metric.
When working with this expression we must be careful to remember that
according to \Deqn{physmet} $\gamma_{\mu\nu}^{\rm TT}=a^2h_{ij}^{\rm TT}$.

The quantity of interest for stochastic background computations is the
energy density in gravitational waves, $\rho_{\rm gw}$. Though it is
rarely stated explicitly, $\rho_{\rm gw}$ is defined with respect to
the proper time of a co-moving observer at rest, i.e.,
\beq
\rho_{\rm gw} \equiv t^\mu t^\nu T_{\mu\nu}^{\rm gw}(t,\mathbf{x}),
\eeq
where $t^\mu=(1,0,0,0)$ in the (background) metric
\beq
ds^2=-dt^2 + a^2(t)d\vec{x}\cdot d\vec{x}.
\eeq
Transforming to conformal coordinates:
\bea
\rho_{\rm gw} &=& t^\mu t^\nu T_{\mu\nu}^{\rm gw}(t,\mathbf{x})
\nonumber
\\
&=& \frac{\eta^\mu\eta^\nu}{a^2(\eta)} T_{\mu\nu}^{\rm
  gw}(\eta,\mathbf{x})
\nonumber
\\
&=& \frac{\eta^\mu\eta^\nu}{32\pi a^2(\eta)}\sum_{i,j}\Bigl\langle\nabla_\mu 
[a^2(\eta)h_{ij}^{\rm TT}]\nabla_\nu  [h_{ij}^{\rm
    TT}/a^2(\eta)]\Bigr\rangle
\nonumber
\\
&=&\frac{1}{32\pi a^2(\eta)}\sum_{i,j}\Bigl\langle\dot{h}^{\rm TT}_{ij}
(\eta,\mathbf{x})\dot{h}^{\rm TT}_{ij}(\eta,\mathbf{x})\Bigr\rangle,
\label{rhogen}
\eea
where $\eta^\mu=(1,0,0,0)$ in conformal coordinates.  Performing the
spatial averaging and using Parseval's theorem we obtain
\beq
\rho_{\rm gw} =\frac{1}{32\pi a(\eta)^2} \frac{1}{V}\sum_{i,j}\int d^3\mathbf{k}\,\, 
\Bigl|\dot{h}_{ij}^{\rm TT}(\eta,\mathbf{k})\Bigr|^2
\label{eqrho}
\eeq
where $V$ is the comoving volume over which the
average is being performed.  We will use this result in both
analytic and computational contexts, which we develop
separately in the next two subsections.

It is important to note that not all modes of the perturbations we
have been discussing are gravitational waves in the
technical sense.  Gravitational waves only exist as such when the
typical wavelength of metric perturbations is much smaller than the
characteristic length scale of the background spacetime, which in the
present context is the Hubble radius.  In other words, waves can only
be observed in situations where a wave zone is well-defined.  In the
current situation, this only happens after the universe has expanded
significantly, e.g., today.  Thus, in order to correctly apply the
formalism developed in the remainder of this section, a (generally
model dependent) procedure must be prescribed for ``transferring'' the
value of $\rho_{\rm gw}$ at the time of creation to today's values
(see Section~\ref{today} for an example).

\subsection{Weinberg-like formula}

Perhaps the most often used tool for analytic calculations of
gravitational radiation is the the formula that appears in 
Weinberg~\cite{137} for the energy in gravitational radiation, $E_{\rm gw}$, as
a function of solid angle\footnote{We have an extra factor of $\pi^2$
relative to the formula in Weinberg's book~\cite{137} due to our Fourier 
transform convention [see our \Deqn{ftcon}].}
\beq
\frac{dE_{\rm gw}}{d\Omega} = \pi^2 \sum_{i,j}
\int_{-\infty}^\infty d\omega\, \omega^2
\bigl| T^{\rm TT}_{ij}(\omega,\mathbf{k}) \bigr|^2.
\label{weinberg}
\eeq
Though the formula is simple its usefulness is
limited to applications in Minkowski space.  It is straightforward to
derive similar expressions for use in radiation- and matter-dominated
(spatially flat) FRW spacetimes.

Imagine that our source, $T_{ij}^{\rm TT}$, is only active for some
finite period of time before ``turning off\/".  Then we are free to take
the upper and lower limits of \Deqn{eqhab} to infinity.  At the same time, we 
can write the source as $T_{ij}^{\rm TT}(\omega,\mathbf{k})$ using the inverse
Fourier transform defined in \Deqn{ftcon}.  Interchanging the $\eta$ and 
$\omega$ integrals then leads to
\beq
h_{ij}^{\rm TT}(\eta,\mathbf{k}) =-i\sqrt{512\pi^3}\frac{\sin({\omega\eta})}{\omega\eta}
\frac{\partial}{\partial\omega}T_{ij}^{\rm TT}(\omega,\mathbf{k}),\label{hijwlf}
\eeq
where the details of the derivation can be found in Appendix \ref{wlfderivation}.
It follows directly that 
\beq
\dot{h}_{ij}^{\rm TT}(\eta,\mathbf{k}) = i\sqrt{512\pi^3}\frac{\sin({\omega\eta})- 
\omega\eta\cos(\omega\eta)}{\omega\eta^2} \frac{\partial}{\partial
 \omega} T_{ij}^{\rm TT}(\omega,\mathbf{k}).\label{dhijwlf}
\eeq
Using these expressions and setting $a(\eta)=\alpha\eta$ in \Deqn{eqrho}, we have 
\bea
\frac{dE_{\rm gw}}{d\Omega}&=&\frac{16\pi^2}{\alpha^2\eta^6} \sum_{i,j}\int_{-\infty}^\infty d\omega\,
\bigl[\sin({\omega\eta}) -  \omega\eta\cos(\omega\eta)\bigr]^2\nonumber\\
&\pheq&\phantom{\frac{16\pi^2}{\eta^6}\int_{-\infty}^\infty d\omega\,}\times\Biggl|
\frac{\partial}{\partial \omega}T_{ij}^{\rm TT}(\omega,\mathbf{k})\Biggr|^2.\label{wlf}
\eea
A similar expression holds for the case of matter-dominated expansion and any 
other simple power law evolution by using the Green's function in \Deqn{gengf}.

\subsection{\label{compmethods}Computational methods}

In many situations, such as the one we will consider in the next
section, the source of gravitational radiation contains a nonlinear
interaction term and we must resort to computer simulations for a
result.  The quantity of interest is given in conformal coordinates by
\beq
\frac{d\rho_{\rm gw}}{d\ln \omega} = 
\frac{\omega^3}{32\pi a^2(\eta)} \frac{1}{V} \sum_{i,j} \int d\Omega\,
\Bigl| \dot{h}_{ij}^{\rm TT}(\eta,\mathbf{k}) \Bigr|^2.
\label{drhodlnkactual}
\eeq
This expression follows directly from \Deqn{eqrho} and is proportional
to $\Omega_{\rm gw}$, the ratio of energy density in
gravitational waves to the energy density required to close the
universe.  In practice, the integral over solid angle requires one to
sum over the entire cubic lattice, which is an ${\cal O}(N^3)$ operation,
where $N$ is the number of lattice points in each direction.

In some situations, to reduce the computational complexity, a trick
introduced in~\cite{Dufaux:2007pt} may be used.  If we assume the
stochastic background is isotropic, we can perform the integral
analytically along any particular direction.  That is, we can write
\beq
\frac{d\rho_{\rm gw}}{d\ln \omega} = \frac{\omega^3}{8 a^2(\eta)} 
\frac{1}{V}\sum_{i,j} 
\Bigl| \dot{h}_{ij}^{\rm TT}(\eta,\omega\mathbf{\hat{k}_p}) \Bigr|^2,
\label{rhodirection}
\eeq
where $\mathbf{\hat{k}_p}$ is a unit vector in the
direction we have chosen.  In practice there are direction
dependent statistical fluctuations but these can be reduced by
averaging \Deqn{rhodirection}
over several directions.  There is a slight subtlety involved: For a
cubic lattice with $N^3$ points, the maximum length (in momentum space) of
any component of $\mathbf k$ is proportional to $N/2$, so that in
three dimensions the maximum possible length of $\mathbf k$ is
$\omega_{\rm max}\propto\sqrt{3}(N/2)$.  However, the length along any
particular direction
is generally shorter than this so the task is to try and maximize both
$\omega_{\rm max}$ and the number of directions with that maximum length
$\omega_{\rm max}$.  Following~\cite{Dufaux:2007pt}, we choose six face
diagonals of the cube.  Namely, the directions that run diagonally
across the $x$--$y$, $x$--$z$, $y$--$z$, $(-x)$--$y$, $(-x)$--$z$ and
$(-y)$--$z$ planes.  This leads to a reasonable reduction in
the effect of statistical fluctuations, with 
$\omega_{\rm max}\propto\sqrt{2}(N/2)$.  Further details of our 
computational methods are given in Appendix~\ref{compdet}.

\section{\label{example}Example: Preheating}

In this section we will consider, as an application of our formalism,
the stochastic background generated from a simple model of preheating
after inflation.  This is an active subject that has been studied with
various techniques, which makes it ideal for validating our methods.
We will start with a brief introduction to the physics of preheating.  
Then we will discuss how one uses the output of the simulation to determine
the spectrum of the stochastic background today before providing a discussion of 
our results.

\subsection{Background}
At the end of inflation the universe is very cold and far from thermal 
equilibrium, which poses a challenge for the the hot big bang scenario.

The earliest attempts at describing reheating after inflation were centered
on the idea that the oscillations of the inflaton about its minimum produce
standard model particles that then interact and bring the universe into thermal
equilibrium~\cite{Linde:1981mu,Albrecht:1982mp,Dolgov:1982th,Abbott:1982hn}.
It was later discovered that this process is preceded by a stage of 
exponential particle production which became known as 
preheating~\cite{Traschen:1990sw,Kofman:1994rk,Kofman:1997yn}.  

We will consider a simple model of preheating in which the inflaton, $\phi$, is 
coupled to some scalar field, $\chi$, according to the Lagrangian
\beq
{\cal L} =\sqrt{-g}\Biggl( \frac{1}{2}\nabla_\mu\phi\nabla^\mu\phi + \frac{1}{2}
\nabla_\mu\chi\nabla^\mu
\chi  - V(\phi,\chi)\Biggr),\label{lagrangian}
\eeq
where $V(\phi,\chi)$ contains both the inflationary potential and the coupling 
between the two fields. In this work we will limit our consideration to 
\beq
V(\phi,\chi) =  \frac{g^2}{2}\phi^2\chi^2 + \frac{\lambda}{4}\phi^4.\label{vpc}
\eeq

The coupling term between $\phi$ and $\chi$ in \Deqn{lagrangian} enters the 
Lagrangian as a spacetime dependent mass term for both fields.  This non-linear 
term can produce an effect know as parametric resonance.  Perhaps the most 
familiar example of parametric resonance is that of a child on a swing at the 
playground. The child, essentially a pendulum, learns fairly quickly if she 
swings her legs back and forth, effectively changing the length of the pendulum
in a time-dependent manner, that she can swing higher and higher, having driven the 
system into resonance.  

\begin{figure}[t]
{\centering
\resizebox*{1\columnwidth}{.22\textheight}
{\includegraphics{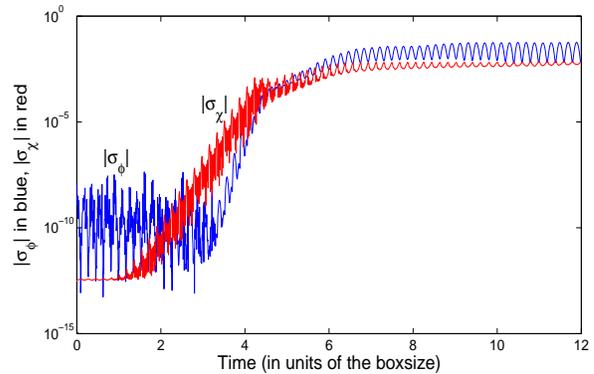}} \par}
\caption{ 
Variances of $\phi$ and $\chi$ fields, as computed by {\sc LatticeEasy}. Note
that these units do not account for the expansion of the universe.
}
\label{fig:variances}
\end{figure}

A more precise description of the dynamics is given by~\cite{Greene:1997fu}, which
is partially summarized in the following.  At the end of inflation, the $\chi$ 
field starts off with zero amplitude and the oscillations of the inflaton about 
its minimum are described in terms of a conformally rescaled field $
\varphi=a\phi$ by the equation 
\beq
\ddot{\varphi} + \lambda\varphi^3 = 0,
\eeq
which has as its solution the Jacobi elliptic cosine function
\beq
\varphi=\frac{x}{\eta\sqrt{\lambda}}\cn\Biggl(x-x_0,\frac{1}{\sqrt{2}}\Biggr),
\eeq
where $x=\eta\phi_0\sqrt{\lambda}$ and $\phi_0$ is the initial value of
the inflaton.  The evolution of the $k$-modes of the 
conformally rescaled $X_k=a\chi_k$ is described by (the Lam\'e equation)
 \beq
 \ddot{X}_k + \Biggl[\frac{k^2}{\lambda\phi_0^2}+\frac{g^2}{\lambda}\cn\Biggl
(x-x_0,\frac{1}{\sqrt{2}}
\Biggr)\Biggr] X_k=0.
 \eeq
This is essentially the harmonic oscillator with a time-dependent
frequency given by the term in square brackets.  The combination
$g^2/\lambda\equiv q$ governs the strength of the time-dependent term
in the frequency and is generally called the resonance parameter.  In
our simulations we typically use $q\sim 100$, which results in
complicated dynamics.  The coupling between $\chi$ and the inflaton 
means that once the oscillations of the inflaton excite parametric
resonance in $\chi$, it in turn brings the inflaton into a stage of
parametric resonance, which continues until this process becomes
inefficient. This is seen quite clearly in
\Dfig{fig:variances}, which shows the variances of each of the
fields as a function of time for one of our simulations (described in
Section~\ref{results}).  Though the variances themselves are of no
particular interest, they provide a simple way to visualize the time
evolution of the full three dimensional dynamics.

\subsection{\label{today}Values today}

As discussed in Section~\ref{gravrad}, we need a prescription for
translating metric perturbations in the early universe into gravitational
radiation today. Our discussion of the transfer function, included for 
completeness, follows~\cite{Dufaux:2007pt,Easther:2007vj}. 
Gravitational energy density scales like $a^{-4}$ so that
\beq
a_{\rm 0}^4\rho_{\rm gw, 0} = a_{\rm e}^4\rho_{\rm gw,e},\label{gwscale}
\eeq
where the 0 and e subscripts denote today and the end of our simulations, 
respectively.  Similarly, if entropy is conserved then radiation energy density 
scales like~\cite{Kolb:1990vq}
\beq
a_{\rm 0}^4g_{\rm 0}^{1/3}\rho_{\rm rad, 0} = a_{\rm e}^4g_{\rm
  e}^{1/3}\rho_{\rm rad,e},
\label{radscale}
\eeq
where $g_{\rm 0}$ and $g_{\rm e}$ are the number of
relativistic species today and at the end time of our simulation (we
take $g_{\rm e}/g_{\rm 0}=100$ for GUT scale inflation)
and the  energy density in radiation at the end of our
simulations is, in the case of radiation-dominated expansion, just the
total energy density at the end of the simulation (i.e., $\rho_{\rm
rad,e}=\rho_{\rm tot,e}$).  The quantity of interest is actually
\beq
\Omega_{\rm gw}h^2 = \frac{1}{\rho_{\rm crit}}\frac{d\rho_{\rm gw}}{d\ln \omega},
\eeq
where $\rho_{\rm crit}=3H_0^2/8\pi$ is the critical density required to close the 
universe and $h$ is a dimensionless factor that accounts for the uncertainty in the
value of the Hubble expansion rate today so that $\Omega_{\rm gw}h^2$ is 
independent of this uncertainty.  We can then use \Deqns{gwscale} and 
(\ref{radscale}) to write
\bea
\Omega_{\rm gw,0}h^2 &=& \Omega_{\rm rad,0}h^2 \Biggl(\frac{g_0}{g_e}\Biggr)^{1/3}
\frac{1}{\rho_{\rm tot, e}}\frac{d\rho_{\rm gw,e}}{d\ln \omega}\nonumber\\
&\approx&\frac{9.3\times 10^{-6}}{\rho_{\rm tot, e}}\frac{d\rho_{\rm gw,e}}{d\ln \omega},
\label{omega1}
\eea
after inserting the appropriate values.  Similarly, we want the physical 
frequency today, $f_{0}=\omega_0/2\pi=\omega_{\rm e}/2\pi a_0$, which is easily 
obtained using \Deqn{radscale}
\bea
f_0 &=& \frac{\omega_{\rm e}}{2\pi a_0}\nonumber\\
    &=& \frac{\omega_{\rm e}}{2\pi a_{\rm e}}\Biggl(\frac{g_0}{g_e}\Biggr)^{1/12} 
        \Biggl(\frac{\rho_{\rm rad,0}}{\rho_{\rm rad,e}}\Biggr)^{1/4}\nonumber\\
    &\approx& \frac{\omega_{\rm e}}{a_{\rm e}\rho_{\rm tot,e}^{1/4}} 
        (4\times 10^{10}{\rm ~Hz}).\label{ftoday}
\eea
The same expressions are found in~\cite{Dufaux:2007pt} and~\cite{Easther:2007vj}.

\subsection{\label{results}Results}

As stated previously, we use {\sc LatticeEasy} for the evolution of
the scalar fields.  Conveniently, the two field model specified 
by \Deqn{lagrangian} is already setup in {\sc LatticeEasy}.  For
our comparison we used the same parameters as many of the simulations
published in~\cite{Dufaux:2007pt} and~\cite{Easther:2007vj}. We took
$\lambda=10^{-14}$ and $g^2=1.2\times 10^{-12}$, which sets the
resonance parameter $q=120$. The initial values of the fields are 
 $\phi_0=0.342M_p$ and $\chi_0=0$, where $M_p$ is the Planck mass. 
This choice of parameters sets $\lambda\phi_0^4/4 
\sim (10^{15} {\rm GeV})^4$, which means inflation happens at the GUT scale.
Additionally, {\sc LatticeEasy} sets initial fluctuations of the fields as described
in~\cite{Felder:2000hq}.
In conformal Planck units we took our box size to be $L=20$, evolved
the box for a time $\eta_f=240$ using timesteps $\Delta\eta=0.01$, and
took $N=256$ points along each of the three axis (the typical size
used in previous publications).  We are able to run this simulation,
including gravitational wave computations, in 18 hours on
a four core 3 GHz machine. The details of our implementation can be
found in Appendix~\ref{compdet}.

\begin{figure}[t]
{\centering
\resizebox*{1\columnwidth}{.22\textheight}
{\includegraphics{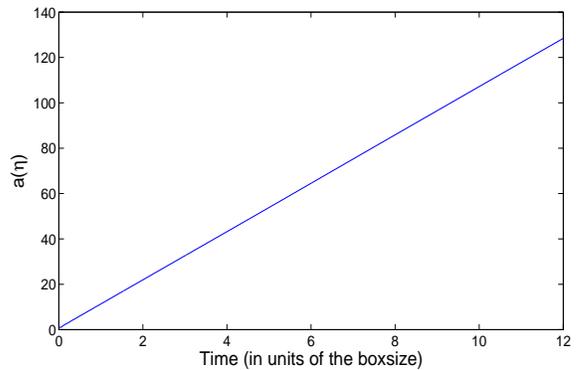}} \par}
\caption{ 
Evolution of the scale factor as a function of time, computed by 
{\sc LatticeEasy}. Note that, to an extremely good approximation, the
evolution is completely radiation-dominated.
}
\label{fig:scale_fac}
\end{figure}

Figure~\ref{fig:scale_fac} shows the self-consistent evolution of the
scale factor as a function of time (the preferred unit for time in all
of our plots is the box size of the simulation).  The straight line
indicates that the expansion is radiation-dominated throughout the
course of the simulation.  More importantly, it validates our use of
the radiation era Green's function.  

To compute the stochastic background spectrum we use \Deqn{omega1}. 
The value of $\rho_{\rm tot, e}$ is computed by {\sc
  LatticeEasy}. To estimate $d\rho_{\rm gw,e}/d\ln \omega$ we use 
\Deqn{rhodirection} computed along six different directions 
as described at the end of Section~\ref{compmethods}. Since the
evolution is very well described by radiation-dominated expansion the values of 
$\dot{h}_{ij}^{\rm TT}(\eta,\omega\mathbf{\hat{k}_p})$ are updated at each
time-step using 
\bea
\dot{h}_{ij}^{\rm TT}(\eta,\omega\mathbf{\hat{k}_p})
&=&\frac{16\pi}{\omega\eta}\int_{\eta_i}^{\eta}d\eta'\,
\eta' \Bigl\{\omega \cos[\omega(\eta-\eta')] 
\nonumber
\\
&\pheq&\phantom{\frac{16\pi}{\omega\eta}\int_{\eta_i}^{\eta}d\eta'}-\frac{1}{\eta} 
\sin[\omega(\eta-\eta')]\Bigr\}\nonumber\\
&\pheq&\phantom{\frac{16\pi}{\omega\eta}\int_{\eta_i}^{\eta}d\eta'}
\times T_{ij}^{\rm TT}(\eta',\omega\mathbf{\hat{k}_p})
\label{eqhabdot}
\eea
which follows from \Deqn{eqhab}. We do this integral using the
rectangle method which appears to be sufficient. To compute $T_{ij}^{\rm
 TT}(\eta',\omega\mathbf{\hat{k}_p})$ at each time-step we construct the
six independent components of the spatial part of the stress energy
tensor and Fourier transform them. Then we project out
everything but the transverse-traceless part using \Deqns{TTT} and
(\ref{proj}) for each of the six $\mathbf{\hat{k}_p}$ we have previously
chosen.  The frequencies are given by \Deqn{ftoday} with 
$\omega_{\rm e} =2\pi i \sqrt{2}/L$, 
where $i=1\ldots N/2$ (the factor of $\sqrt{2}$ comes from our choice
of $\mathbf{\hat{k}_p}$ along the face diagonals of the simulation box; see
Section~\ref{compmethods} for details).

\begin{figure}[t]
{\centering
\resizebox*{1\columnwidth}{.22\textheight}
{\includegraphics{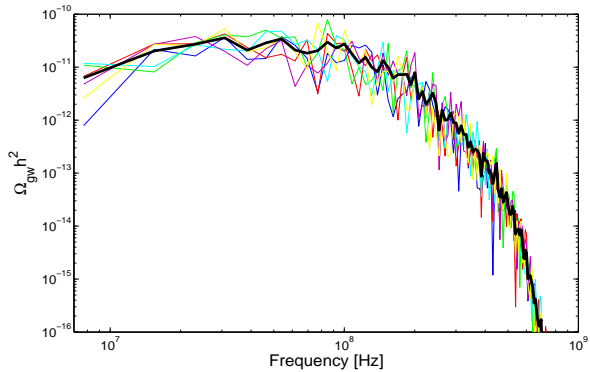}} \par}
\caption{ The spectrum of the stochastic background today, $\Omega_{{\rm
      gw}}h^2$, computed along six directions on the lattice (thin
  colored lines), and the average (thick black line).}
\label{fig:omega}
\end{figure}

\Dfig{fig:omega} shows the result of this procedure.  We plot
$\Omega_{\rm gw}h^2$ as a function of frequency (both are the values
today).  The thin colored lines are the results along each of the six
directions in momentum space and the thick black line is the average.

\begin{figure}[t]
{\centering
\resizebox*{1\columnwidth}{.22\textheight}
{\includegraphics{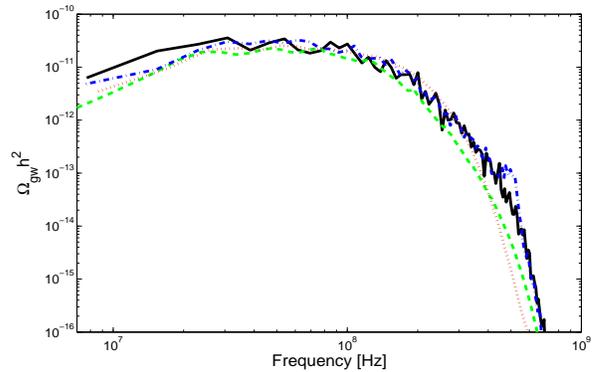}} \par}
\caption{ Comparison of $\Omega_{\rm gw}$ today as computed by the method 
described in this paper (solid black line), as well as the curves published
in~\cite{Dufaux:2007pt} (dash-dotted blue line), \cite{Easther:2007vj} (dashed 
green line) and~\cite{GarciaBellido:2007af} (dotted red line).  Note the good
agreement across methods.}
\label{fig:omega_comparison}
\end{figure}

Next we present a comparison of our results with those obtained by other groups
using different methods.  The authors of~\cite{Dufaux:2007pt}, \cite{Easther:2007vj}, 
and~\cite{GarciaBellido:2007af} kindly provided us with the results of their 
simulations.  The (solid) black curve is the result of our simulation (the same as in 
\Dfig{fig:omega}). The (dash-dotted) blue
curve is produced from the data published in~\cite{Dufaux:2007pt}\footnote{These 
authors provided us with $\Omega_{\rm gw}$ along each axis in addition to the
six directions described in Section~\ref{compmethods}.  Only the latter was 
used in \Dfig{fig:omega_comparison}}. 
The (dashed) green curve shows the simulation results
published in~\cite{Easther:2007vj} for higher resolution simulation with $N=512$ points
in each direction. Finally, the (dotted) red curve is 
from~\cite{GarciaBellido:2007af} for a simulation with $N=128$\footnote{These 
authors provided us with $\Omega_{\rm gw}/\rho_{\rm tot,e}$ and $k$ at the end
of the simulation.  \Dfig{fig:omega_comparison} displays the result of applying the 
transfer functions described in Section~\ref{today} to their data.} The results 
across all methods are in very good agreement. 

It is worth pointing out that the Green's function introduced
in~\cite{Dufaux:2007pt} is an approximation for general kinds of
expansion, but for radiation-dominated expansion it is in fact exact.
Looking at Eq.~(12) of~\cite{Dufaux:2007pt} we see that it is the same
as our \Deqn{eqhab} (their metric perturbation is re-scaled by a
factor of $a(\eta)$).  However, they work in the small wavelength 
approximation and perform an additional average over a single period of
oscillation, which presumably accounts for the slight differences 
between the (dash-dotted) blue and (solid) black curves.

Although the results methods agree quite well with one another, there is a 
general issue common to all simulations: The fact
that $\eta_f=12L$.  This, by itself, is not a problem, but {\sc LatticeEasy}
uses periodic boundary conditions.  This means that excitations of the scalar 
fields leave and re-enter the simulation 
box multiple times, setting up correlations in the field values 
not present in the 
early universe. The magnitude and importance of this effect is unclear.
The problem can be avoided by setting the final time to 
(at most) half of the light-crossing time of the box, i.e., $\eta_f\leq L/2$.
Keeping the same final time, $\eta_f=240$, we would set $L=480$, which would 
require $24^3$ times as many points on the lattice to keep the current 
resolution.  This, in turn would increase the memory required for the simulation 
by a similar factor.  Alternatively, a glance at~\Dfig{fig:variances} suggests
that gravitational wave production is essentially complete by $\eta=120$ (this
can be verified directly by monitoring the evolution in time of 
$\Omega_{\rm gw}$).  A simulation with $\eta_f=120$ and $L=240$ would only
require $12^3$ times as many points to maintain the current resolution.  We 
leave this investigation for future work.

Finally, an obvious question that arises is what to do in situations where there is 
significant gravitational wave production during times when the expansion of the
universe is not purely radiation- or matter-dominated.   While each situation
requires its own analysis, we can, generally speaking, combine our method with
that of~\cite{Dufaux:2007pt}.  That is, we can divide the calculation of 
$\Omega_{\rm gw}$ into periods when the evolution is essentially a simple power
law and use the exact Green's functions given by \Deqn{gengf} during those periods.  In 
the periods in-between, where the evolution is either more complicated or 
changing rapidly, we can use the approximate expressions in~\cite{Dufaux:2007pt}.

\section{\label{conclusions}Conclusions}
In this paper we have presented a method for computing the stochastic background
of gravitational waves due to (classical) sources in a spatially flat FRW 
background.  Our method centers around ``evolving" the metric perturbation using the 
exact Green's functions for radiation- and matter-dominated expansion.  We 
developed these results into a Weinberg-like formula, applicable in expanding
spacetimes, for use in situations where the dynamics of the source is known or
can be approximated.  Additionally, we used the same results to present a 
computational framework applicable to scalar field sources with complicated 
dynamics.  Because our framework relies on Green's functions for 
the evolution of the metric perturbation, it is robust and simple and
we avoid the numerical difficulties associated with more complicated evolution
schemes.  Furthermore, we have shown that the 
results of using this framework to compute the stochastic background produced 
in the $\lambda\phi^4$ model of preheating agree quite well with those appearing 
in the literature.

While the frequency ranges and sensitivities of the gravitational waves from the 
model of preheating studied here make it unlikely to be detected in the near 
future, there remain several other options for the production of gravitational 
waves that could be observable soon.  Aside from the sources discussed in 
Section~\ref{intro} that may be observable to Advanced LIGO, phase transitions 
taking place at the electroweak scale could lead to a signal detectable by 
LISA~\cite{Grojean:2006bp}.  We plan to investigate this in more detail using
the methods introduced here.

\begin{acknowledgments} 
We would like to thank John T. ``Tom'' Giblin, Jr.  and Jolien
Creighton for illuminating discussions. We would also like to thank
Juan Garcia-Bellido, Jeff Dufaux, Richard Easther, Gary Felder, Dani 
Figueroa, Lev Kofman and Alfonso Sastre for useful communications. We are
further grateful to Gary Felder and Igor Tkachev for developing {\sc
LatticeEasy} and making it publicly available. LP is supported by NSF 
PHY-0503366 and the Research Growth Initiative at the University of 
Wisconsin-Milwaukee.  
\end{acknowledgments}

\appendix
\section{\label{mpct}Metric perturbations in cosmological time}
 In this appendix we derive expressions for the metric perturbation in the 
commonly used coordinates
 \beq
 ds^2=-dt^2+a^2(t)(\delta_{ij}+h_{ij}^{\rm TT})dx^idx^j,\label{cosmotmp}
\eeq
where $h_{ij}^{\rm TT}=h_{ij}^{\rm TT}(t,\mathbf{x})$ satisfies Eqs.~(\ref{gc1}--\ref{gc3}).  
One could, in principle, perform a coordinate transformation 
on the expressions in conformal coordinates, but we find it more convenient to 
simply re-derive the results.  In these coordinates the perturbed Einstein 
equations take the form (after performing a spatial Fourier transform):
\beq
h_{ij}^{''\, \rm TT}+3\frac{a'(t)}{a(t)}h_{ij}^{'\, \rm TT}+\frac{\omega^2}{a^2(t)}h_{ij}^{\rm 
TT}=16\pi \frac{T_{ij}^{\rm 
TT}}{a^2(t)}.
\eeq
Focusing once again on scale factor evolution of the form
\beq
a(t)=\alpha t^n,
\eeq
for constant alpha, leads to the Green's function solution
\bea
h_{ij}^{\rm TT}(t,\mathbf{k}) &=& 16\pi\int_{t_i}^{t_<}dt'\,\frac{\pi\Bigl[t^{(1-3n)}t^{\prime\, (1-n)}\Bigr]^{1/2}}{2\alpha^2(n-1)}
\Bigl[J_\beta(z)Y_\beta(z')\nonumber\\
&\pheq&\phantom{16\pi\int_{t_i}^{t_<}dt'\,}
-J_\beta(z')Y_\beta(z)\Bigr]
 T_{ij}^{\rm TT}(t',\mathbf{k}),\nonumber\\
\label{gfct}
\eea
where $J_\beta(z)$ and $Y_\beta(z)$ are Bessel functions of the first and second 
kind, $t_<\equiv \min(t,t_f)$  and
\bea
\beta&=&\frac{1-3n}{2(n-1)},\\
z&=&\frac{\omega t^{1-n}}{\alpha(n-1)},\\
z'&=&\frac{\omega t^{\prime\, 1-n}}{\alpha(n-1)}.
\eea
Note that the factors in the denominators of the preceeding expressions prevent 
them from being valid when $n=1$.  The metric perturbation for $a(t)=\alpha t$ 
is given in terms of $\gamma=\sqrt{1-(\omega/\alpha)^2}$ by
\bea
h_{ij}^{\rm TT}(t,\mathbf{k}) &=& 16\pi\int_{t_i}^{t_<}dt'\,\frac{t'}{2\alpha \gamma t}
\Biggl[t^{\gamma}t^{\prime\,-\gamma}
-t^{-\gamma}t^{\prime\,\gamma}\Biggr]
T_{ij}^{\rm TT}(t',\mathbf{k}).\nonumber\\
\eea   
Particularly interesting cases of \Deqn{gfct} are radiation- and 
matter-dominated expansion which are characterized by
\beq
a(t)=\alpha t^{1/2}\quad {\rm and}\quad a(t)=\alpha t^{2/3},\nonumber
\eeq
respectively.  The corresponding metric perturbations are given by
\beq
h_{ij}^{\rm TT}(t,\mathbf{k}) = \frac{16\pi}{\alpha\omega t^{1/2}}\int_{t_i}^{t_<}dt'
\sin\Biggl[\frac{2\omega}{\alpha}(t^{1/2}-t'^
{1/2})\Biggr] T_{ij}^{\rm TT}(t',
\mathbf{k}),
 \eeq
 for radiation-dominated expansion and 
 \begin{widetext}
 \beq
 h_{ij}^{\rm TT}(t,\mathbf{k}) = \frac{16\pi}{9\alpha\omega^3t}\int_{t_i}^{t_<}dt'\, t'^{-1/3}\Biggl\{\Biggl[1
+\frac{9\omega^2}{\alpha^2}(tt')^{1/3}\Biggr] \sin\Biggl[\frac{3\omega}{\alpha}(t^
{1/3}-t'^{1/3})\Biggr]-\frac{3\omega}{\alpha}(t^{1/3}-t'^{1/3}) \cos\Biggl[\frac{3\omega}{\alpha}(t^{1/3}-t'^{1/3})\Biggr]\Biggr\} T_{ij}^{\rm TT}(t',
\mathbf{k}),
 \eeq
 \end{widetext}
for matter-dominated expansion.

\section{\label{wlfderivation}Details of the calculation of the Weinberg-like formula}
In this appendix we provide the details of the calculation leading to 
\Deqn{hijwlf}, from which the Weinberg-like formula in \Deqn{wlf} follows 
trivially.  As mentioned previously, we begin with \Deqn{eqhab} and take the 
upper and lower limits of the integral to infinity while taking the inverse
Fourier transform of the source.  At the same time we expand the sine into 
exponentials and make use of the identity
\beq
\eta' = \lim_{\zeta\to 0} -i\frac{\partial}{\partial\zeta} e^{i\zeta\eta'}.
\eeq 
Interchanging the order of the $\eta$ and $\omega$ integrals then leads to 
\Deqn{hijwlf}. More specifically 
\begin{widetext}
\bea
h_{ij}^{\rm TT}(\eta,\mathbf{k}) &=&
\frac{16\pi}{\omega\eta}\int_{-\infty}^{\infty}
d\eta'\,\eta'\sin[\omega(\eta-\eta')]T_{ij}^{\rm TT}(\eta',\mathbf{k})
\nonumber
\\
&=& -\frac{8\pi i}{\omega\eta}\Biggl(-i\frac{\partial}
{\partial\zeta}\Biggr) \int_{-\infty}^{\infty} d\eta'\,
e^{i\zeta\eta'}[e^{i\omega(\eta-\eta')}-e^{-i\omega(\eta-\eta')}]
\int_{-\infty}^{\infty}
\frac{d\omega'}{(2\pi)^{1/2}}\,e^{-i\omega'\eta'}
T_{ij}^{\rm TT}(\omega',\mathbf{k})
\nonumber
\\
&=& -\frac{8\pi }{\omega\eta}\frac{\partial}{\partial\zeta}\int_{-\infty}^{\infty}
\frac{d\omega'}{(2\pi)^{1/2}}\,
\int_{-\infty}^{\infty} d\eta'\, 
[e^{i\omega\eta}e^{-i\eta'(\omega'+\omega-\zeta)}
- e^{-i\omega\eta}e^{-i\eta'(\omega'-\omega-\zeta)}]T_{ij}^{\rm TT}(\omega',\mathbf{k}) 
\nonumber
\\
&=&  -\frac{8\pi }{\omega\eta}\frac{\partial}{\partial\zeta}\int_{-\infty}^{\infty}
{d\omega'}{(2\pi)^{1/2}}\, 
\Biggl\{e^{i\omega\eta}\delta(\omega'+\omega-\zeta)
- e^{-i\omega\eta}\delta(\omega'-\omega-\zeta)\Biggr\}
T_{ij}^{\rm TT}(\omega',\mathbf{k}) 
\nonumber
\\
&=&
-\frac{\sqrt{128\pi^3}}{\omega\eta}\Biggl[{e^{i\omega\eta}}\frac{\partial}{\partial
  \zeta}
T_{ij}^{\rm TT}(-\omega+\zeta,\mathbf{k})-{e^{-i\omega\eta}}\frac{\partial}
{\partial \zeta}T_{ij}^{\rm TT}(\omega+\zeta,\mathbf{k})\Bigr)\Biggr]\nonumber\\
&=&-\frac{\sqrt{128\pi^3}}{\omega\eta}\Biggl[{e^{i\omega\eta}}\frac{\partial}{\partial
  \omega}
T_{ij}^{\rm TT}(\omega,\mathbf{k})-{e^{-i\omega\eta}}\frac{\partial}
{\partial \omega}T_{ij}^{\rm TT}(\omega,\mathbf{k})\Bigr)\Biggr]\nonumber\\
&=&-i\sqrt{512\pi^3}\frac{\sin({\omega\eta})}{\omega\eta}\frac{\partial}{\partial
  \omega}
T_{ij}^{\rm TT}(\omega,\mathbf{k}),
\eea
\end{widetext}
where we suppressed the limit symbol throughout, integrated by parts in the 
fifth line and took $\zeta\to 0$ in the sixth line.
From here it is straightforward to compute the derivative with respect to $\eta$
and use the expression for the energy density in gravitational radiation given 
by \Deqn{eqrho} to obtain the Weinberg-like formula (quoted here for 
completeness):
\begin{widetext}
\beq
\frac{dE_{\rm gw}}{d\Omega}= \frac{16\pi^2}{\alpha^2\eta^6}\sum_{i,j}\int_{-\infty}^\infty d\omega\,
\bigl[\sin({\omega\eta}) -  \omega\eta\cos(\omega\eta)\bigr]^2\Biggl|
\frac{\partial}{\partial \omega}T_{ij}^{\rm TT}(\omega,\mathbf{k})\Biggr|^2.
\eeq
\end{widetext}

\section{\label{compdet}Computational details}
Our code for computing the stochastic background interfaces directly with 
{\sc LatticeEasy}.  We ran our simulations on a four core 3 GHz machine.
To take advantage of th multi-processing capabilities available to us 
we used OpenMP~\cite{openmp} as implemented in gcc version 4.2.3.
The loops over the lattice necessary to evaluate the six independent 
components of $T_{ij}$ were split into four threads, each evaluated on one core.
The most computationally expensive tasks we perform are the Fourier
transforms of $T_{ij}$ [see \Deqns{eqhab} and (\ref{eqhabmattdom})].  
There are six Fourier transforms to perform at each timestep.  
For this task we enlist the help of the Fastest Fourier Transform in the 
West (FFTW)~\cite{FFTW05}.  Although FFTW has its own multi-processor capabilities, 
we find that we can get $\sim 10\%$ better performance 
using OpenMP for the task.  By trial and error we found six threads, one
for each Fourier transform, to be the fastest option. 
Note that we have not modified {\sc LatticeEasy}.  With these 
improvements the gravitational wave computation takes roughly the same amount
of time as the field evolution. The simulation described in Section 
~\ref{results} takes 18 hours to complete.

\bibliography{refs}

\end{document}